\documentclass[letterpaper, 10 pt, conference]{ieeeconf}%
\usepackage{subfigure}
\usepackage{bm}
\usepackage{amssymb}
\usepackage{epsfig}
\usepackage{amsmath}
\usepackage{graphicx}
\usepackage{epstopdf}
\usepackage{amsfonts}%
\usepackage{indentfirst}
\usepackage{eufrak}%
\newtheorem{problem}{\textbf{Problem}}
\newtheorem{definition}{\rm\textbf{Definition}}
\newtheorem{theorem}{\rm\textbf{Theorem}}

\newtheorem{remark}{\rm\textbf{Remark}}
\providecommand{\U}[1]{\protect\rule{.1in}{.1in}}
\IEEEoverridecommandlockouts
\begin{document}

\title{{\LARGE \textbf{Control Barrier Functions for Systems with Multiple Control Inputs}}}
\author{Wei Xiao, Christos G. Cassandras, Calin A. Belta and Daniela Rus \thanks{This work was
		supported in part by NSF under grants ECCS-1931600, DMS-1664644, CNS-1645681, IIS-1723995, and IIS-2024606, by ARPAE under grant DE-AR0001282, by AFOSR under grant FA9550-19-1-0158, by the United States AFRL and Air Force Artificial Intelligence Accelerator under Cooperative Agreement Number FA8750-19-2-1000, and by the MathWorks. The views and conclusions contained in this document are those of the authors and should not be interpreted as representing the official policies, either expressed or implied, of the United States Air Force or the U.S. Government. The U.S. Government is authorized to reproduce and distribute reprints for Government purposes notwithstanding any copyright notation herein. }\thanks{W. Xiao and D. Rus are with the Computer Science and Artificial Intelligence Lab, Massachusetts Institute of Technology \texttt{{\small weixy@mit.edu, rus@csail.mit.edu}}} \thanks{C. G. Cassandras and C. Belta are
with the Division of Systems Engineering and Center for Information and
Systems Engineering, Boston University, Brookline, MA, 02446, USA
\texttt{{\small \{cgc, cbelta\}@bu.edu}}}}
\maketitle

\begin{abstract}
%It has been shown that optimizing quadratic costs while stabilizing affine control systems to desired (sets of) states subject to state and control constraints can be reduced to a sequence of Quadratic Programs (QPs) by using Control Barrier Functions (CBFs) and Control Lyapunov Functions (CLFs). 
Control Barrier Functions (CBFs) are becoming popular tools in guaranteeing safety for nonlinear systems and constraints, and they can reduce a constrained optimal control problem into a sequence of Quadratic Programs (QPs) for affine control systems. The recently proposed High Order Control Barrier Functions (HOCBFs) work for arbitrary relative degree constraints. One of the challenges in a HOCBF is to address the relative degree problem when a system has multiple control inputs, i.e., the relative degree could be defined with respect to different components of the control vector. This paper proposes two methods for HOCBFs to deal with systems with multiple control inputs: a general integral control method and a method which is simpler but limited to specific classes of physical systems. When control bounds are involved, the feasibility of the above mentioned QPs can also be significantly improved with the proposed methods. We illustrate our approaches on a unicyle model with two control inputs, and compare the two proposed methods to demonstrate their effectiveness and performance.

\end{abstract}

\thispagestyle{empty} \pagestyle{empty}

%%%%%%%%%%%%%%%%%%%%%%%%%%%%%%%%%%%%%%%%%%%%%%%%%%%%%%%%%%%%%%%%%%%%%%%%%%%%%%%%

%%%%%%%%%%%%%%%%%%%%%%%%%%%%%%%%%%%%%%%%%%%%%%%%%%%%%%%%%%%%%%%%%%%%%%%%%%%%%%%%
\section{Introduction}
\label{sec:intro}

Barrier functions (BFs) are Lyapunov-like functions \cite{Tee2009}, 
\cite{Wieland2007}, whose use can be traced back to optimization problems
\cite{Boyd2004}. More recently, they have been employed to prove set
invariance \cite{Aubin2009}, \cite{Prajna2007}, \cite{Wisniewski2013} and for
multi-objective control \cite{Panagou2013}. In \cite{Tee2009}, it was proved
that if a BF for a given set satisfies Lyapunov-like conditions, then the set
is forward invariant. 
Control BFs (CBFs) are extensions of BFs for control systems that are used to
map a constraint defined over system states onto a constraint on the control
input. Recently, it has been shown that, to stabilize an affine control system
while optimizing a quadratic cost and satisfying state and control
constraints, CBFs can be combined with Control Lyapunov Functions (CLFs)
\cite{Sontag1983}, \cite{Artstein1983}, \cite{Freeman1996} to form a sequence of
quadratic programs (QPs) \cite{Galloway2013}, \cite{Ames2017}, 
\cite{Glotfelter2017}.

The CBFs from \cite{Ames2017} and \cite{Glotfelter2017} work for constraints
that have relative degree one with respect to the system dynamics. A general form \cite{Nguyen2016}
for arbitrarily high relative degree constraints, termed exponential CBF, employs input-output
linearization and finds a pole placement controller with negative poles. The high order
CBF (HOCBF) proposed in \cite{Xiao2019} is simpler and more general than the
exponential CBF \cite{Nguyen2016}.  However, when we have a system with {\it multiple} control inputs, an additional problem arises in a HOCBF when considering the relative degree of a safety constraint. In other words, the relative degree could be defined with respect to different components of the control vector. How to make a desired subset or all the control components show up in the HOCBF constraint remains an open problem.

In order to deal with systems with multiple control inputs, this paper proposes to guarantee constraint satisfaction through HOCBFs with desired control components of the control vector using two methods: integral control (for general systems) and a simpler constraint transformation-based approach (for a specific class of systems). In the integral control method, we define auxiliary dynamics for the control components that are differentiated in the corresponding HOCBF constraint. Then, we consider integral CBFs \cite{Ames2020} to guarantee constraint satisfaction while having the desired control components in the HOCBF constraint. In the transformation method, we transform a safety constraint into a new constraint that forces all the desired control components to show up in the corresponding HOCBF constraint. Unlike the integral control method, the latter, although simpler, works only for a certain class of systems, which are described later in the paper. The satisfaction of this new constraint implies the satisfaction of the original constraint. When control bounds are involved, the feasibility of the QPs can be significantly improved if all the control components show up in the HOCBF constraint.  We demonstrate the effectiveness of our methods on a unicycle model with two controls, and compare their relative performance.

%This paper is structured as follows. In Sec. \ref{sec:pre}, we provide background and preliminary results on HOCBFs and CLFs. We propose two different methods to guarantee constraint satisfaction for systems with multiple control inputs in Sec. \ref{sec:mcbf}. A unicycle control problem is formulated in Sec. \ref{sec:prob}, followed by simulation results in Sec. \ref{sec:sim}. We conclude with final remarks in Sec. \ref{sec:conc}.

\section{Preliminaries}
\label{sec:pre}

 \begin{definition}
 	\label{def:classk} (\textit{Class $\mathcal{K}$ function} \cite{Khalil2002}) A
 	continuous function $\alpha:[0,a)\rightarrow[0,\infty), a > 0$ is said to
 	belong to class $\mathcal{K}$ if it is strictly increasing and $\alpha(0)=0$. A continuous function $\beta:\mathbb{R}\rightarrow\mathbb{R}$ is said to belong to extended class $\mathcal{K}$ if it is strictly increasing and $\beta(0)=0$.
 \end{definition}

 Consider an affine control system of the form
 \begin{equation}
 \dot{\bm{x}}=f(\bm x)+g(\bm x)\bm u \label{eqn:affine}%
 \end{equation}
 where $\bm x\in\mathbb{R}^{n}$, $f:\mathbb{R}^{n}\rightarrow\mathbb{R}^{n}$
 and $g:\mathbb{R}^{n}\rightarrow\mathbb{R}^{n\times q}$ are {locally}
 Lipschitz, and $\bm u\in U\subset\mathbb{R}^{q}$ with the control constraint set $U$
 defined as
 \begin{equation}
 U:=\{\bm u\in\mathbb{R}^{q}:\bm u_{min}\leq\bm u\leq\bm u_{max}\}.
 \label{eqn:control}%
 \end{equation}
 $\bm u_{min},\bm u_{max}\in\mathbb{R}^{q}$ and the inequalities are
  componentwise.
 
 \begin{definition}
 	\label{def:forwardinv} A set $C\subset\mathbb{R}^{n}$ is forward invariant for
 	system (\ref{eqn:affine}) if its solutions for some $\bm u\in U$ starting at any $\bm x(0) \in C$ satisfy $\bm x(t)\in C,$ $\forall t\geq0$.
 \end{definition}
 
 \begin{definition}
 	\label{def:relative} (\textit{Relative degree}) The relative degree of a
 	(sufficiently many times) differentiable function $b:\mathbb{R}^{n}%
 	\rightarrow\mathbb{R}$ with respect to system (\ref{eqn:affine}) is the number
 	of times it needs to be differentiated along its dynamics until the control
 	$\bm u$ explicitly shows in the corresponding derivative.
 \end{definition}

 The above definition works for a system with a single control input. In Sec. \ref{sec:mcbf}, we will provide an extension for systems with multiple control inputs. 
 Since function $b$ is used to define a (safety) constraint $b(\bm
 x)\geq0$, we will also refer to the relative degree of $b$ as the relative
 degree of the constraint. For a constraint $b(\bm x)\geq0$ with relative
 degree $m$, $b:\mathbb{R}^{n}\rightarrow\mathbb{R}$, and $\psi_{0}(\bm
 x):=b(\bm x)$, we define a sequence of functions $\psi_{i}:\mathbb{R}%
 ^{n}\rightarrow\mathbb{R},i\in\{1,\dots,m\}$:
 \begin{equation}
 \begin{aligned} \psi_i(\bm x) := \dot \psi_{i-1}(\bm x) + \alpha_i(\psi_{i-1}(\bm x)),\quad i\in\{1,\dots,m\}, \end{aligned} \label{eqn:functions}%
 \end{equation}
 where $\alpha_{i}(\cdot),i\in\{1,\dots,m\}$ denotes a $(m-i)^{th}$ order
 differentiable class $\mathcal{K}$ function.
 
 We further define a sequence of sets $C_{i}, i\in\{1,\dots,m\}$ associated
 with (\ref{eqn:functions}) in the form:
 \begin{equation}
 \label{eqn:sets}\begin{aligned} C_i := \{\bm x \in \mathbb{R}^n: \psi_{i-1}(\bm x) \geq 0\}, \quad i\in\{1,\dots,m\}. \end{aligned}
 \end{equation}

 \begin{definition}
 	\label{def:hocbf} (\textit{High Order Control Barrier Function (HOCBF)}
 	\cite{Xiao2019}) Let $C_{1}, \dots, C_{m}$ be defined by (\ref{eqn:sets}%
 	) and $\psi_{1}(\bm x), \dots, \psi_{m}(\bm x)$ be defined by
 	(\ref{eqn:functions}). A function $b: \mathbb{R}^{n}\rightarrow\mathbb{R}$ is
 	a High Order Control Barrier Function (HOCBF) of relative degree $m$ for
 	system (\ref{eqn:affine}) if there exist $(m-i)^{th}$ order differentiable
 	class $\mathcal{K}$ functions $\alpha_{i},i\in\{1,\dots,m-1\}$ and a class
 	$\mathcal{K}$ function $\alpha_{m}$ such that
 	\begin{equation}
 	\label{eqn:constraint}\begin{aligned} 
 	\sup_{\bm u\in U}[L_f^{m}b(\bm x) + [L_gL_f^{m-1}b(\bm x)]\bm u \!+\! O(b(\bm x)) \\+ \alpha_m(\psi_{m-1}(\bm x))] \geq 0, \end{aligned}
 	\end{equation}
 	for all $\bm x\in C_{1} \cap,\dots, \cap C_{m}$. 
 	In
 	(\ref{eqn:constraint}), $L_{f}^{m}$ ($L_{g}$) denotes the Lie derivative along
 	$f$ ($g$) $m$ (one) times, and $O(\cdot) = \sum_{i = 1}^{m-1}L_f^i(\alpha_{m-i}\circ\psi_{m-i-1})(\bm x)$. %{Further, $b(\bm x)$ is such that $L_gL_f^{m-1}b(\bm x)\ne 0$ on the boundary of the set $C_{1} \cap,\dots, \cap C_{m}$.}
 \end{definition}
 
 The HOCBF is a general form of the relative degree one CBF
 \cite{Glotfelter2017} \cite{Ames2017}, (setting $m=1$ reduces the HOCBF to
 the common CBF form in \cite{Glotfelter2017}, and it is also a more general form of the exponential CBF
 \cite{Nguyen2016}. {Note that we can define $\alpha_i(\cdot), i\in\{1,\dots, m\}$ in Def. \ref{def:hocbf} to be extended class $\mathcal{K}$ functions to ensure robustness to perturbations \cite{Ames2017}. }
 
 \begin{theorem}
 	\label{thm:hocbf} (\cite{Xiao2019}) Given a HOCBF $b(\bm x)$ from Def.
 	\ref{def:hocbf} with the associated sets $C_{1}, \dots, C_{m}$ defined
 	by (\ref{eqn:sets}), if $\bm x(0) \in C_{1} \cap,\dots,\cap C_{m}$,
 	then any Lipschitz continuous controller $\bm u(t)$ that satisfies
 	the constraint in (\ref{eqn:constraint}), $\forall t\geq0$ renders $C_{1}\cap,\dots,
 	\cap C_{m}$ forward invariant for system (\ref{eqn:affine}).
 \end{theorem}
 
% \begin{definition}
% 	\label{def:clf} (\textit{Control Lyapunov function (CLF)} \cite{Aaron2012}) A
% 	continuously differentiable function $V: \mathbb{R}^{n}\rightarrow\mathbb{R}$
% 	is an exponentially stabilizing control Lyapunov function (CLF) for system
% 	(\ref{eqn:affine}) if there exist constants $c_{1} >0, c_{2}>0, c_{3}>0$ such
% 	that for $\forall\bm x\in\mathbb{R}^{n}$ for which $c_{1}||\bm x||^{2} \leq V(\bm x)
% 	\leq c_{2} ||\bm x||^{2},$ we have
% 	\begin{equation}
% 	\label{eqn:clf}\underset{u\in U}{\inf} \lbrack L_{f}V(\bm x)+L_{g}V(\bm x)
% 	\bm u + c_{3}V(\bm x)\rbrack\leq0.
% 	\end{equation}
 %\end{definition}
 
 Many existing works \cite{Ames2017}, \cite{Nguyen2016}
 combine CBFs for systems with relative degree one with quadratic costs to form
 optimization problems. Time is discretized and an optimization problem with
 constraints given by the CBFs (inequalities of the form (\ref{eqn:constraint}%
 )) is solved at each time step. Note that these
 constraints are linear in control since the state value is fixed at the
 beginning of the interval, therefore, each optimization problem is a quadratic
 program (QP) if the cost is quadratic in the control. The optimal control obtained by solving each QP is applied at
 the current time step and held constant for the whole interval. Since the aforementioned QPs are myopically solved pointwise, these QPs can easily become
 infeasible when control
 bounds are also involved.  We have recently shown that this problem may be overcome by finding sufficient conditions that can guarantee the feasibility of the QP at each time step \cite{Xiao2021}. This CBF method works well for systems with single input. However, for systems with multiple controls, it is not clear how to choose the relative degree $m$ in a HOCBF, which we address in this paper.

\section{Systems with Multiple Inputs}
\label{sec:mcbf}

In this section, we consider how we may guarantee (safety) constraint satisfaction for systems with multiple control inputs with HOCBFs while having a desired subset or all of the control components show up in the corresponding HOCBF constraint (\ref{eqn:constraint}). 

Suppose there is a safety requirement $b(\bm x)\geq 0$ for system (\ref{eqn:affine}). If we enforce this constraint using a HOCBF, we first need to determine the relative degree of $b(\bm x)$. As system (\ref{eqn:affine}) may have multiple control inputs, we may consider the relative degree as the minimum number of times that we  differentiate $b(\bm x)$ along the dynamics (\ref{eqn:affine}) until any component of the control vector shows up in the corresponding derivative. However, this may reduce the system performance (such as only limited control components can be used to guarantee constraint satisfaction); an example is given below.

For instance, to make an autonomous vehicle satisfy a safety constraint with respect to a preceding vehicle using HOCBFs, we may require the ego vehicle to follow the preceding vehicle or overtake it. In the former case, steering wheel control is not desired to show up in the HOCBF constraint (\ref{eqn:constraint}) as the ego vehicle fails to follow the preceding vehicle. However, in the latter case, we wish that both the acceleration control and steering wheel control show up in the HOCBF constraint (\ref{eqn:constraint}). This can improve the mobility of an autonomous vehicle compared to the former case, as well as significantly improve the feasibility of the resulting HOCBF-based QPs when conrol bounds are involved.

In this section, we focus on the case where we wish {\it all} the components of the control vector to show up in the HOCBF constraint (\ref{eqn:constraint}). We can always fix the control (e.g., setting it to 0) of undesired control components in the HOCBF constraint (\ref{eqn:constraint}) if we wish to guarantee safety using {\it some} desired control components.
In order to achieve this, we define the relative degree set $S\subset \mathbb{N}$ of a function $b:\mathbb{R}^n\rightarrow\mathbb{R}$ as follows:
\begin{definition} \label{def:relatives}
	({\it Relative degree set})
	The relative degree set $S\subset \mathbb{N}$ of a function $b:\mathbb{R}^n \rightarrow \mathbb{R}$ with respect to system (\ref{eqn:affine}) is defined by the set of numbers of times we need to differentiate $b$ along system (\ref{eqn:affine}) until each component of the control vector $\bm u$ first shows in the corresponding derivative. 
\end{definition}

 This definition is illustrated in the following example.

\subsection{Motivating Example}
\label{sec:moti}

Consider a simplified unicycle model  of the form: %\cite{Solea2007}
\begin{equation} \label{eqn:unicycle}
\dot x = v\cos\theta,\; \dot y = v\sin\theta,\; \dot v = \frac{u_2}{M},\; \dot \theta = \phi,\; \dot \phi = u_1,
%\underbrace{\left[
%\begin{array}
%[c]{c}%
%\dot{x}\\
%\dot{y}\\
%\dot v\\
%\dot \theta\\
%\dot \phi\\
%\end{array}
%\right]}_{\dot{\bm x}}  = \underbrace{\left[
%\begin{array}
%[c]{c}%
%v\cos\theta\\
%v\sin\theta\\
%0\\
%\phi\\
%0
%\end{array}
%\right]}_{f(\bm x)} + \underbrace{\left[\begin{array}{cc}  
%	0 & 0\\
%	0 & 0\\
%	0 & \frac{1}{M}\\
%	0 & 0\\
%	1 & 0
%	\end{array} \right]}_{g(\bm x)}\underbrace{\left[\begin{array}{c}  
%	u_{1}\\
%	u_{2}
%	\end{array} \right]}_{\bm u} 
\end{equation}
where $(x, y)\in\mathbb{R}^2$ denotes the 2-D location of the system, $v\in\mathbb{R}$ denotes the linear speed, $\theta\in\mathbb{R}$ is the heading angle, $\phi\in\mathbb{R}$ denotes the rotation speed, $M > 0$ is the mass of the system, and $u_1 \in\mathbb{R}, u_2 \in\mathbb{R}$ stand for the angular acceleration and driven force (control inputs), respectively.

Suppose we have a constraint for system state:
\begin{equation} \label{eqn:safety}
\sqrt{(x - x_0)^2 + (y - y_0)^2}\geq r,
\end{equation}
where $(x_0, y_0)\in \mathbb{R}^2$, and $r> 0$.
With $b(\bm x) := \sqrt{(x - x_0)^2 + (y - y_0)^2} - r$, we can see that the relative degree of $b$ with respect to $u_1$ is 3, and the relative degree of $b$ with respect to $u_2$ is 2. Therefore, the relative degree set of $b(\bm x)$ is $S=\{3, 2\}$.

When defining a HOCBF (taking class $\mathcal{K}$ functions $\alpha_1, \alpha_2$ as linear functions) for the constraint (\ref{eqn:safety}) with respect to $u_2$, i.e., the relative degree $m = 2$ in Def. \ref{def:hocbf}, we have
\begin{equation} \label{eqn:hocbfconstr1}
L_f^2b(\bm x) + L_{g_2}L_{f}b(\bm x)u_2 + 2L_fb(\bm x) + b(\bm x) \geq 0.
\end{equation}
where $g_2 = [0, 0, \frac{1}{M}, 0, 0]^T$ is the second column of $g(\bm x)$ in (\ref{eqn:unicycle}).

We can also define a HOCBF (taking class $\mathcal{K}$ functions $\alpha_1, \alpha_2, \alpha_3$ as linear functions) for the constraint (\ref{eqn:safety}) with respect to $u_1$, i.e., the relative degree $m = 3$ in Def. \ref{def:hocbf}, and we have
\begin{equation} \label{eqn:hocbfconstr2}
\begin{aligned}
L_f^3b(\bm x) \!+\! L_{g_1}L_{f}^2b(\bm x)u_1 \!+\! L_f[L_{g_2}L_{f}b(\bm x)]u_2 \! +\! L_{g_2}L_{f}b(\bm x)\dot u_2 \\+ 3L_f^2b(\bm x) + 3L_{g_2}L_fb(\bm x)u_2 + 3L_fb(\bm x) + b(\bm x) \geq 0,
\end{aligned}
\end{equation}
 where $g_1 = [0, 0, 0, 0, 1]^T$ is the first column of $g(\bm x)$ in (\ref{eqn:unicycle}). Note that the derivative of $u_2$ is included in the above.

There is only one control input in the HOCBF constraint (\ref{eqn:hocbfconstr1}). The safety constraint (\ref{eqn:safety}) can only be guaranteed using $u_2$ (not $u_1$) and the CBF-CLF based QP feasibility is also impaired when control bounds as in (\ref{eqn:control}) are present. In other words, a robot can only use the linear deceleration to avoid the obstacle specified by the constraint (\ref{eqn:safety}). On the other hand, (\ref{eqn:hocbfconstr2}) includes the derivative of $u_2$, and this introduces an additional problem (i.e., how to choose $\dot u_2$) in the HOCBF-based QP.
In order to address these issues, we propose two approaches as shown in the following sections.

%In a general setting, consider a safety requirement $b(\bm x) \geq 0$ for system (\ref{eqn:affine}), and we wish to enforce it with a HOCBF. We assume that the cardinality of the relative degree set $S$ for $b(\bm x)$ is greater than one, otherwise, we just have a HOCBF that includes all the components of $\bm u$.

\subsection{Integral HOCBFs}
We begin by introducing some notations that will facilitate the analysis that follows. There may be some control components of $\bm u$ that are differentiated at least once in the HOCBF constraint, and we define the index set of those differentiated control components as $S_d$. The cardinality of $S_d$ is denoted by $N_d \in\mathbb{N}, N_d < q$ (recall that $q$ is the dimension of $\bm u$). Let $\bm u_n$ denote the control vector that includes only those control components whose derivatives are never present in the corresponding HOCBF constraint (e.g., $\bm u_n = u_1$ in (\ref{eqn:hocbfconstr2})), and let $U_n$ denote the control constraint set (defined as in (\ref{eqn:control})) corresponding to $\bm u_n$. Let $g_n:\mathbb{R}^n\rightarrow\mathbb{R}^{n\times (q - N_d)}$ denote a matrix that is composed of the columns of the matrix $g(\bm x)$ in (\ref{eqn:affine}) corresponding to each control component in $\bm u_n$.

Let $\overline{m}$ denote the maximum relative degree of $b(\bm x)\geq 0$ with respect to (\ref{eqn:affine}), i.e., $\overline{m} = \max_{k\in S} k$. Then, we define  $b(\bm x)$ to be a HOCBF with relative degree $\overline{m}$.  In order to deal with the derivatives of the control components, we define auxiliary dynamics for each $u_j, j\in S_d$ that is differentiated $m_j$ times, $1\leq m_j < \overline{m}$:
\begin{equation}\label{eqn:aux}
\dot {\bm u}_j = f_j(\bm u_j) + g_j(\bm u_j) \nu_j,
\end{equation}
where $\bm u_j = (u_{j,1}, u_{j,2}, \dots, u_{j,m_j})\in \mathbb{R}^{m_j}$ is the corresponding auxiliary state, and $u_{j,1} = u_j$, $f_j:\mathbb{R}^{m_j}\rightarrow \mathbb{R}^{m_j}, g_j:\mathbb{R}^{m_j}\rightarrow \mathbb{R}^{m_j}$, and $\nu_j\in\mathbb{R}$ is a new control for the auxiliary dynamics (\ref{eqn:aux}) corresponding to $u_j$. The relative degree of $u_j$ (now a state variable) with respect to (\ref{eqn:aux}) is $m_j$, and $\bm u_j$ can be initialized to any vector as long as $u_j$ strictly satisfies its control bound in (\ref{eqn:control}). Although $f_j, g_j$ may be arbitrarily selected, for simplicity, we may define (\ref{eqn:aux}) in linear form, and initialize $u_{j,k}, k\in\{2,\dots, m_j\}$ to 0. Further, let $\bm \nu\in\mathbb{R}^{N_d}$ be the concatenation of $\nu_j, \forall j\in S_d$, and let $\bm u_a$ be the concatenation of $\bm u_j, \forall j\in S_d$. Thus, $u_j, j\in S_d$, unlike $\bm u_n$, contains all control components whose derivatives appear at least once in the  HOCBF constraint. 

Combining the auxiliary dynamics (\ref{eqn:aux}), we get the HOCBF constraint (relative degree $\overline{m}$) enforcing $b(\bm x)\geq 0$:
\begin{equation} \label{eqn:icbf}
\begin{aligned}
L_f^{\overline{m}}b(\bm x) + [L_{g_n}L_f^{\overline{m}-1}b(\bm x)]\bm u_n \!+\! R(b(\bm x), \bm u_a, \bm \nu) \\+ \alpha_{\overline{m}}(\psi_{\overline{m}-1}(\bm x, \bm u_a)) \geq 0,
\end{aligned}
\end{equation}
where $R(\cdot)$ is defined similar to $O(\cdot)$ as in (\ref{eqn:constraint}), but also includes the derivatives of $u_j, \forall j\in S_d$, i.e., $u_j^{(1)},\dots, u_j^{(m_j)}$, and $u_j^{(m_j)}$ denotes the $m_j^{th}$ derivative of $u_j$.

In order to apply the CBF-based QP approach to guarantee (safety) constraint satisfaction, we can take $\bm u_n, \nu_j, \forall j\in S_d$ instead of $\bm u$ as the decision variables in the QP. After solving the QP, we obtain the optimal $\bm u_n, \nu_j, \forall j\in S_d$ for each time interval, and the controls $u_j, j\in S_d$ are obtained by solving (\ref{eqn:aux}). Since this is done by integration, we call this {\it integral control}. We refer to the resulting HOCBF in (\ref{eqn:icbf}) as an integral HOCBF (iHOCBF), and it is a class of integral control barrier functions \cite{Ames2020}.

As in (\ref{eqn:control}), we have control bounds for each $u_j, j\in S_d$:
\begin{equation} \label{eqn:control_sing}
u_{j,min} \leq u_j \leq u_{j,max},
\end{equation}
where $u_{j, min}\in\mathbb{R}, u_{j,max}\in \mathbb{R}$ denote the minimum and maximum control bounds, respectively.
In order to guarantee the above control bound (\ref{eqn:control_sing}) under the auxiliary dynamics (\ref{eqn:aux}), we define two HOCBFs for each $u_j, j\in S_d$ to map the bound from $u_j$ to $\nu_j$. Letting $b_{j,min}(\bm u_j) = u_j - u_{j,min}$ and $b_{j,max}(\bm u_j) = u_{j,max} - u_j$. 
%we have the following two HOCBF constraints:
%\begin{equation} \label{eqn:bound1}
%\begin{aligned}
%L_{f_j}^{m_j}b_{j,min}(\bm u_j) \!+\! [L_{g_j}L_{f_j}^{m_j-1}b_{j,min}(\bm u_j)] \nu_j \!+\! O(b_{j,min}(\bm u_j)) \\+ \alpha_{m_j}(\psi_{m_j-1}(\bm u_j)) \geq 0, j\in S_d
%\end{aligned}
%\end{equation}
%\begin{equation}\label{eqn:bound2}
%\begin{aligned}
%L_{f_j}^{m_j}b_{j,max}(\bm u_j) \!+\! [L_{g_j}L_{f_j}^{m_j-1}b_{j,max}(\bm u_j)] \nu_j \!+\! O(b_{j,max}(\bm u_j)) \\+ \alpha_{m_j}(\psi_{m_j-1}(\bm u_j)) \geq 0, j\in S_d
%\end{aligned}
%\end{equation}
We can then define the set of the auxiliary control $\bm \nu$ that enforces (\ref{eqn:control_sing}):
\begin{equation} \label{eqn:contrl}
\begin{aligned}
U_a(\bm u_a) = \{\bm \nu\in\mathbb{R}^{N_d}: L_{f_j}^{m_j}b_{j,min}(\bm u_j) + O(b_{j,min}(\bm u_j)) \\+ [L_{g_j}L_{f_j}^{m_j-1}b_{j,min}(\bm u_j)] \nu_j  + \alpha_{m_j}(\psi_{m_j-1}(\bm u_j)) \geq 0,\\ L_{f_j}^{m_j}b_{j,max}(\bm u_j) + [L_{g_j}L_{f_j}^{m_j-1}b_{j,max}(\bm u_j)] \nu_j \\+ O(b_{j,max}(\bm u_j)) + \alpha_{m_j}(\psi_{m_j-1}(\bm u_j)) \geq 0, \forall j\in S_d\}.
\end{aligned}
\end{equation}
Through the equations above, the states of the auxiliary dynamics (\ref{eqn:aux}) are strictly bounded for each $\nu_j, j\in S_d$. This property is captured by the invariant sets defined as in (\ref{eqn:sets}):

\begin{definition} \label{def:hocbfm2}
	({\it Integral HOCBFs}) Let $C_1, C_2,\dots, C_{\overline{m}}$ be defined as in (\ref{eqn:sets}) and $\psi_0(\bm x, \bm u_a), \psi_1(\bm x, \bm u_a),\dots, \psi_{\overline{m}}(\bm x, \bm u_a)$ be defined as in (\ref{eqn:functions}). A function $b: \mathbb{R}^n\rightarrow \mathbb{R}$ is an integral HOCBF (iHOCBF) of relative degree $\overline{m}$ for system (\ref{eqn:affine}) if there exist differentiable class $\mathcal{K}$ functions $\alpha_1,\alpha_2,\dots, \alpha_{\overline{m}}$ and auxiliary dynamics (\ref{eqn:aux}) such that
	\begin{equation}\label{eqn:constraintm2}
	\begin{aligned}
	\sup_{\bm u_n\in U_n, \bm \nu\in U_a(\bm u_a)}[L_f^{\overline{m}}b(\bm x) + [L_{g_n}L_f^{\overline{m}-1}b(\bm x)]\bm u_n \\+ R(b(\bm x), \bm u_a, \bm\nu) + \alpha_{\overline{m}}(\psi_{\overline{m}-1}(\bm x, \bm u_a))] \geq 0,
	\end{aligned}
	\end{equation}
	for all $(\bm x, \bm u_a)\in C_1 \cap C_2\cap,\dots, \cap C_{\overline{m}}$. 
\end{definition}

Note that $u_j, \forall j\in S_d$ and their derivatives become state variables with the auxiliary dynamics (\ref{eqn:aux}). All control inputs $\bm u_n, \bm \nu$ are in linear forms if we fix $\bm x, \bm u_j, \forall j\in S_d$.  Similar to Thm. \ref{thm:hocbf}, we also have the following theorem:
\begin{theorem} \label{thm:hocbfm2}
	Given an iHOCBF $b(\bm x)$ from Def. \ref{def:hocbfm2} with the associated sets $C_1, C_2,\dots, C_{\overline{m}}$ defined as in (\ref{eqn:sets}), if $(\bm x(t_0), \bm u_a(t_0)) \in C_1 \cap C_2\cap,\dots,\cap C_{\overline{m}}$, then any continuously differentiable controller $(\bm u_n(t), \bm \nu(t))\in U_n\times U_a(\bm u_a), \forall t\geq t_0$ that satisfies the constraint in (\ref{eqn:constraintm2})
	renders the set
	$C_1\cap C_2\cap,\dots, \cap C_{\overline{m}}$ forward invariant for systems (\ref{eqn:affine}), (\ref{eqn:aux}). Moreover, the control $\bm u_n$ and integral controls $u_j, \forall j\in S_d$ render $C_1$ forward invariant for system (\ref{eqn:affine}).
\end{theorem}

\textbf{Proof:} Combining dynamics (\ref{eqn:affine}) and the auxiliary dynamics (\ref{eqn:aux}), we still have an affine control system with $\bm u_n, \bm \nu$ as the control inputs. In other words, the combined dynamics are in the form:
\begin{equation}
\dot{\bm y} = F(\bm y) + G(\bm y) \bm u_y
\end{equation}
where $\bm y = (\bm x, \bm u_a), \bm u_y = (\bm u_n, \bm \nu), F:\mathbb{R}^{n + \sum_{j\in S_d}m_j}\rightarrow\mathbb{R}^{n + \sum_{j\in S_d}m_j}, G:\mathbb{R}^{n + \sum_{j\in S_d}m_j}\rightarrow \mathbb{R}^{(n + \sum_{j\in S_d}m_j)\times q}$.

Considering the above dynamics, the HOCBF constraint (\ref{eqn:constraintm2}) is equivalent to:
\begin{equation}\label{eqn:constraintm22}
\begin{aligned}
\sup_{\bm u_y\in U_n\times U_a(\bm u_a)}[L_F^{\overline{m}}b(\bm y) + [L_{G}L_F^{\overline{m}-1}b(\bm y)]\bm u_y \\+ O(b(\bm y)) + \alpha_{\overline{m}}(\psi_{\overline{m}-1}(\bm y))] \geq 0,
\end{aligned}
\end{equation}
where $b(\bm y) =b(\bm x)$.

By Thm. \ref{thm:hocbf}, we have that the set
$C_1\cap C_2\cap,\dots, \cap C_{\overline{m}}$ is forward invariant for systems (\ref{eqn:affine}), (\ref{eqn:aux}). As $u_j, \forall j\in S_d$ are obtained through the integration of (\ref{eqn:aux}), each $u_j$ is differentiable,  then we have that the set $C_1$ is forward invariant for system (\ref{eqn:affine}). $\qquad\qquad\qquad\qquad\qquad\qquad\;\;\;\;\;\blacksquare$

\textbf{Example revisited.} For the motivating example, the minimum and maximum relative degrees of (\ref{eqn:safety}) are 2 and 3, respectively. If we define a HOCBF with relative degree 3 to enforce (\ref{eqn:safety}), we will have the derivative of $u_2$ in the corresponding HOCBF constraint. Following the above process, we may define the following simple auxiliary dynamics for $u_2$: $\dot u_2 = \nu$. Then, the HOCBF (taking class $\mathcal{K}$ functions $\alpha_1, \alpha_2, \alpha_3$ as linear functions) constraint (\ref{eqn:constraintm2}) for (\ref{eqn:safety}) is
\begin{equation} \label{eqn:ihocbf}
\begin{aligned}
L_f^3b(\bm x(t)) + L_{g_1}L_f^2b(\bm x(t))u_1(t) +  L_f[L_{g_2}L_fb(\bm x(t))]u_2(t) \\+ L_{g_2}L_fb(\bm x(t)) \nu + 3L_f^2b(\bm x(t)) + 3L_{g_2}L_fb(\bm x(t))u_2(t)\\+ 3L_fb(\bm x(t)) + b(\bm x(t)) \geq 0,
\end{aligned}
\end{equation}
where $u_2$ in the above is a state variable instead of a decision variable (control) in the QP. The resulting $u_2$ applied to system (\ref{eqn:unicycle}) is obtained through the integration of the auxiliary dynamics $\dot u_2 = \nu$. Meanwhile, the control bound for $\nu$ is obtained through the two CBFs in (\ref{eqn:contrl}).

\subsection{HOCBFs based on Constraint Transformation}
\begin{figure}[thpb]
	\vspace{-2mm}
	\centering
	$\hspace{-6mm}$\includegraphics[scale=0.12]{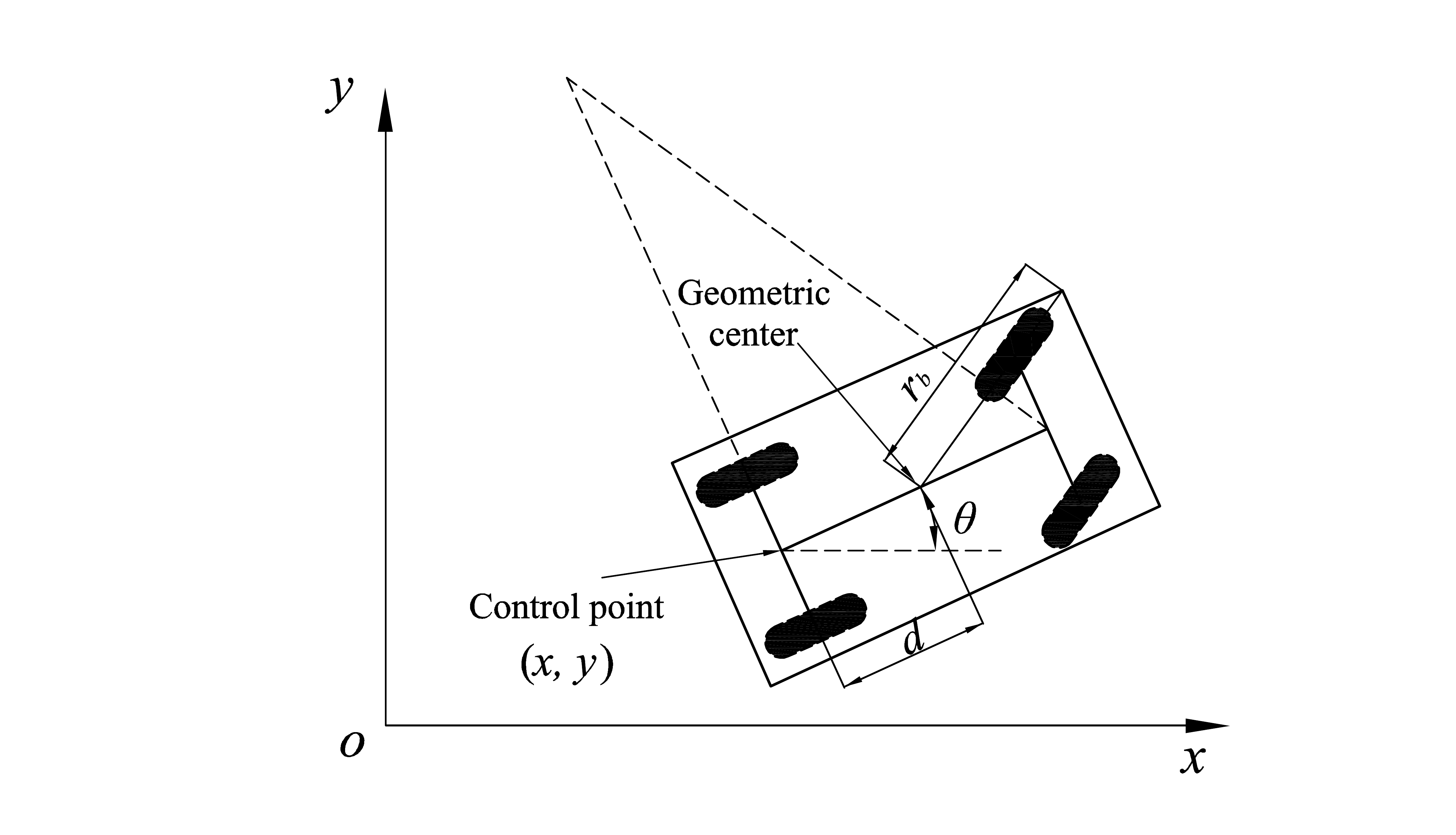}
	\vspace{-4mm}
	\caption{The geometric structure of a vehicle. }	
	\label{fig:vehicle}
	\vspace{-2mm}
\end{figure}

The integral control method in the last section involves the derivatives of some control components. However, this can be actually avoided if we can determine a transformation of the constraint into a new one which has a unique relative degree with respect to (\ref{eqn:affine}) and such that the associated HOCBF includes all the components of $\bm u$. Clearly, this is not always possible and requires some extra structure in the system. In particular, the transformation approach we propose in this section targets a special class of systems corresponding to physical objects (e.g., robotic systems) with well-defined geometric structures (e.g., a vehicle has a rectangular footprint, as shown in Fig. \ref{fig:vehicle}). This approach can be extended to other types of systems, as long as we can find a reasonable constraint transformation that avoids the presence of control input derivatives as in (\ref{eqn:hocbfconstr2}).

A control point (origin) on a system is the point for which we define dynamics (\ref{eqn:affine}), as the one shown in Fig. \ref{fig:vehicle}. Since the system we consider has a geometric boundary, we need to make sure that any point on this system never violates safety constraints. This can be (conservatively) achieved by considering the safety requirement on the system geometric center, i.e., the geometric center should stay away from unsafe sets with an additional distance that is determined by the geometry of the system. For the vehicle example in Fig. \ref{fig:vehicle}, the additional safe distance with respect to the center is $r_b > 0$, where $r_b$ is the radius of the circle circumscribing the geometric boundary.

In such systems, we map safety constraints from the control point to the system geometric center (or other point that plays a similar role). The original constraint is given by $b(\bm x) \geq 0$ (on the control point),  which may be (geometrically) mapped to a new one (on the geometric center): 
\begin{equation}
b_{T}(\bm x) \geq 0,
\end{equation}
 where $b_{T}:\mathbb{R}^n\rightarrow \mathbb{R}$. The satisfaction of the above constraint implies the satisfaction of the original constraint $b(\bm x)\geq 0$. An example of finding such a transformation function is given at the end of this section. The specification of $b_{T}$ is such that it has a unique relative degree with respect to (\ref{eqn:affine}) and all the control components of $\bm u$ show in the corresponding HOCBF constraint. Note that this approach may not work for some systems, in which case we can turn to the integral control approach introduced in the last section.

Letting $m_t$ denote the (unique) relative degree of $b_{T}$, we have a definition of an extra mixed relative degree CBF as:
\vspace{2ex}
\begin{definition} \label{def:hocbfm3}
	({\it HOCBFs based on constraint transformation}) Let $C_1,\dots, C_{m_t}$ be defined as in (\ref{eqn:sets}) and $\psi_0(\bm x), \dots, \psi_{m_t}(\bm x)$ be defined as in (\ref{eqn:functions}), a function $b: \mathbb{R}^n\rightarrow \mathbb{R}$ is a HOCBF with transformation of relative degree $m_t$ for system (\ref{eqn:affine}) if there exist differentiable class $\mathcal{K}$ functions $\alpha_1,\alpha_2,\dots, \alpha_{m_t}$, a transformation function $b_{T}:\mathbb{R}^n\rightarrow \mathbb{R}$ whose relative degree is uniquely determined by $m_t$, and $b_{T}(\bm x)\geq 0$ implies $b(\bm x)\geq 0$ such that
	\begin{equation}\label{eqn:constraintm3}
	\begin{aligned}
	L_f^{m_t}b_{T}(\bm x) + L_gL_f^{m_t-1}b_{T}(\bm x)\bm u + O(b_{T}(\bm x)) \\+ \alpha_{m_t}(\psi_{m_t-1}(\bm x)) \geq 0
	\end{aligned}
	\end{equation}
	for all $\bm x\in C_1 \cap C_2\cap,\dots, \cap C_{m_t}$. In the above equation, both $C_1, C_2,\dots, C_{m_t}$ and $\psi_0(\bm x), \psi_1(\bm x),\dots, \psi_{m_t}(\bm x)$ are defined by $b_{T}$ instead of $b$. $O(\cdot)$ denotes the remaining Lie derivatives along $f$ with degree less than $m_t$ (omitted for simplicity).
\end{definition}

Since $b_{T}$ has a unique relative degree $m_t$, all control inputs $u_1,u_2,\dots,u_q$ show up in (\ref{eqn:constraintm3}), and the constraint (\ref{eqn:constraintm3}) is linear on control inputs. Similar to Thm. \ref{thm:hocbf}, we also have the following theorem:
\begin{theorem} \label{thm:hocbfm3}
	Given a HOCBF $b(\bm x)$ from Def. \ref{def:hocbfm3} with the associated sets $C_1, C_2,\dots, C_{m_t}$ defined as in (\ref{eqn:sets}), if $\bm x(t_0) \in C_1 \cap C_2\cap,\dots,\cap C_{m_t}$, then any Lipschitz continuous controller $\bm u(t)\in U, \forall t\geq t_0$ that satisfies the constraint in (\ref{eqn:constraintm3})
	guarantees $b(\bm x)\geq 0$ for system (\ref{eqn:affine}).
\end{theorem}

	The proof of the above theorem is simple as we have $C_1 \cap C_2\cap,\dots,\cap C_{m_t}$ is forward invariant by Thm. \ref{thm:hocbf}, i.e., $b_{T}(\bm x) \geq 0$ is guaranteed. The constraint transformation method might be conservative as $b_T(\bm x)\geq 0$ might not imply $b(\bm x)\geq 0$. Note that since $b_{T}(\bm x) \geq 0$ implies $b(\bm x) \geq 0$, we also have that $b(\bm x) \geq 0$ is always satisfied. An example is given below.

\textbf{Example revisited.} Consider a vehicle with the geometry structure shown in Fig. \ref{fig:vehicle}, and the dynamics are as in (\ref{eqn:unicycle}). The location of the control point of the model is $(x, y)\in\mathbb{R}^2$, and its geometric center location is given by $(x + d\cos(\theta), y + d\sin(\theta))$, where $d > 0$. If we consider the safety constraint (\ref{eqn:safety}) which does not take into account any geometric structure on the control point, then we need to define $b(\bm x):= \sqrt{(x - x_0)^2 + (y - y_0)^2} - r - r_v\geq 0$ in order to avoid collision between the robot and the obstacle described by the safety constraint, where $r_v$ is determined by $d, r_b$ in Fig. \ref{fig:vehicle} (i.e., the distance between the control point and the corner). However, if we consider the safety constraint (\ref{eqn:safety}) on the geometric center, then we have a transformation function $b_{T}(\bm x) = b(x + d\cos(\theta), y + d\sin(\theta)) + r_v - r_b$ that guarantees collision avoidance. The relative degrees of $b_{T}$ with respect to (\ref{eqn:unicycle}) corresponding to $u_1$ and $u_2$ are both 2, thus, the HOCBF (taking class $\mathcal{K}$ functions $\alpha_1, \alpha_2$ as linear functions) constraint (\ref{eqn:constraintm3}) which in this case is
\begin{equation} \label{eqn:thocbf}
\begin{aligned}
L_f^2b_{T}(\bm x) + L_gL_fb_{T}(\bm x)
\left[\begin{array}{c}  
u_1\\
u_2
\end{array} \right] 
+ 2L_fb_{T}(\bm x) + b_{T}(\bm x) \geq 0,
\end{aligned}
\end{equation}

\begin{remark} \label{remark:compare}
	(Comparison between the integral control and transformation methods) The integral control method works for general systems that have multiple control inputs, but the resulting HOCBFs have higher relative degree compared with the transformation method. Higher relative degree may cause additional difficulties when we consider the feasibility of a HOCBF in an unknown environment as there are more class $\mathcal{K}$ functions. The transformation method only works for specific physical systems for which we can find a transformation function from the original safety constraint with the property that the relative degrees of all the control components are the same. The computational cost in the transformation method is lower than the integral control method as the HOCBFs usually have lower relative degree. Therefore, in practice, we always seek to apply the constraint transformation method first.
\end{remark}

\subsection{Optimal Control for Systems with Multiple Inputs}

Consider an optimal control problem for system (\ref{eqn:affine}) in which there are multiple inputs with the cost defined as:
\begin{equation}\label{eqn:cost}
J(\bm u(t)) = \int_{t_0}^{t_f}\mathcal{C}(||\bm u(t)||)dt,
\end{equation}
where $||\cdot||$ denotes the 2-norm of a vector, and  $\mathcal{C}(\cdot)$ is a strictly increasing function. Assume a constraint $b(\bm x) \geq 0$ has to be satisfied by system (\ref{eqn:affine}). Then the control input $\bm u$ should satisfy the HOCBF constraint (\ref{eqn:constraintm2}) or (\ref{eqn:constraintm3}).

If convergence to a given state is required in addition to optimality and safety, then, as 
in \cite{Ames2017}, the HOCBF can be combined with a CLF. We discretize time and formulate a cost (\ref{eqn:cost}) subject to the HOCBF constraint (\ref{eqn:constraintm2}) or (\ref{eqn:constraintm3}) and the CLF constraint at each time step. With the optimal control input $\bm u$ obtained from the QP at each time step, we update the system dynamics (\ref{eqn:affine}), and the procedure is repeated. Then, the safety constraint $b(\bm x(t)) \geq 0$ is satisfied for (\ref{eqn:affine}), $\forall t\in[t_0, t_f]$, and the system performance and QP feasibility are improved since all the control components show up in the corresponding HOCBF constraint. This will be illustrated in the following case study.

%\section{Robot Control}
\section{Case Study}
\label{sec:prob}

We consider a robot with the unicycle model as in (\ref{eqn:unicycle}). The robot has a circular shape with radius $r_b > 0$, and its control point is displaced from the geometric center by $d$, where $0<d<r_b$ (see Fig. \ref{fig:vehicle}).

%\subsection{Problem Formulation}
\textbf{Objective}: We consider a cost in the form:
	$
	J(\bm u(t)) \!=\! \int_{t_0}^{t_f} \left[u_1^2(t) + u_2^2(t)\right] dt + p||\bm x_p(t_f) - \bm X||,
	$
where $p > 0, \bm x_p = (x,y)$ and $\bm X\in\mathbb{R}^2$ is a terminal position.

\begin{figure}[thpb]
	\vspace{-2mm}
	\centering
	\includegraphics[scale=0.35]{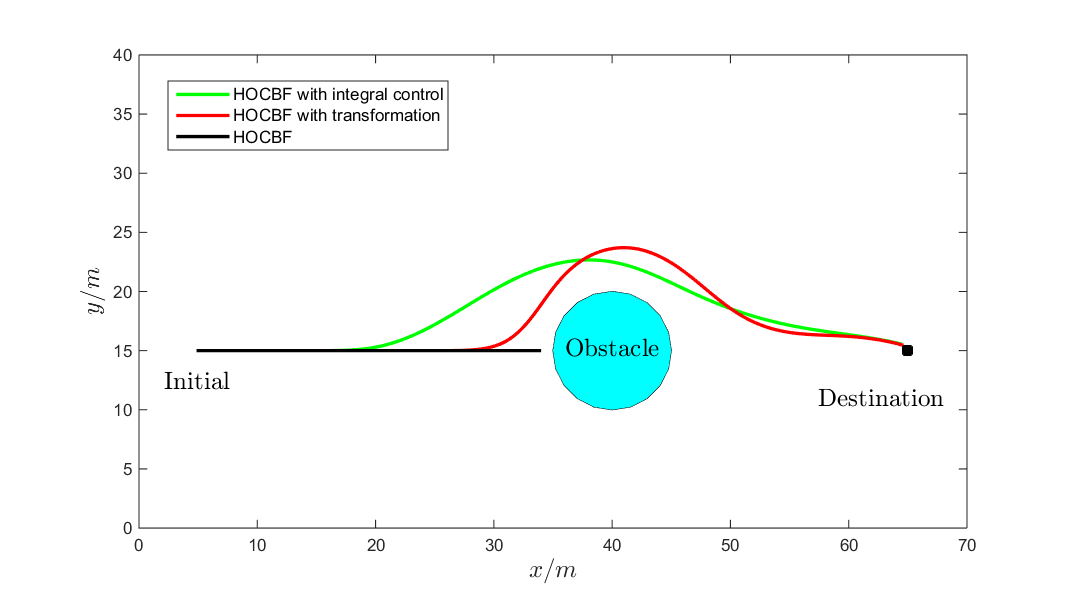}
	\vspace{-4mm}
	\caption{Robot trajectories under different forms of HOCBFs. }	
	\label{fig:traj}
\end{figure}

\textbf{Constraint 1} (Safety constraint): The robot should avoid collision with a circular obstacle (see Fig. \ref{fig:traj}), i.e., it should satisfy a constraint imposed on the control point $(x,y)$:
\begin{equation} \label{eqn:safety2}
\sqrt{(x - x_0)^2 + (y - y_0)^2}\geq r + r_b + d,
\end{equation}

\textbf{Constraint 2} Robot Limitations: The state and control limitations are defined as:
$
v_{min} \leq v \leq v_{max}, \qquad
\phi_{min} \leq \phi \leq \phi_{max}, 
u_{1,min} \leq u_1 \leq u_{1,max},\quad u_{2,min} \leq u_2 \leq u_{2,max},
$
where $v_{min}\geq 0, v_{max}> 0$,  $\phi_{min}\geq 0, \phi_{max}> 0$, $u_{1,min} < 0, u_{1,max}> 0$, and $u_{2,min}< 0, u_{2,max}> 0$.

\begin{problem}\label{problem}
	Determine a control law to minimize Objective 1 subject to Constraints 1, 2, for the robot governed by dynamics (\ref{eqn:unicycle}).
\end{problem}

%The relative degrees of (\ref{eqn:safety2}) with respect to (\ref{eqn:unicycle}) are 2 and 3 corresponding to $u_2$ and $u_1$, respectively, and the relative degree set $S = \{3,2\}$. 
We use iHOCBFs to implement (\ref{eqn:safety2}), and use a CLF to enforce the desired terminal state in Obective 1.
We conducted simulations in MATLAB to compare the effectiveness and performance of the proposed HOCBFs for systems with multiple control inputs. The simulation parameters are $v_0 = 5m/s, \Delta t = 0.1s, (x_d, y_d) = (65,15), r = 5m, r_b = 1m, d = 0.5m, M = 1650kg, \phi_{max} = -\phi_{min} = 0.6981rad/s, v_{max} = 5m/s, v_{min} = 0, u_{1,max} = -u_{1,min} = 0.3491rad/s^2, u_{2,max} = -u_{2, min} = 3M.$ The robot initial state vector is $\bm x(t_0) = (5m, 15m, 0, 0, 0)$. 

%\begin{table}[!htb]
%	\caption{Simulation parameters for problem \ref{problem}}
%	\label{tab:param}
%	\centering
%	\begin{tabular}{|c||c||c||c||c||c|}\hline
%		Name     & value     & unit & Name     & value     & unit \\\hline\hline
%		$p_{ang}, p_{spe}$ & 1  & unitless   &$v_0$     & 5       & $m/s$    \\\hline
%		$\varepsilon$     & 10 & unitless & $\Delta t$     & 0.1      & $s$   \\\hline
%		$x_d$ & 65 &  $m$ & $y_d$ & 15 &  $m$  \\\hline
%		$r$ & 5 &  $m$ & $r_b (d)$ & 1 (0.5) &  $m$  \\\hline
%		$M$ & 1650 &  $kg$ & $g$ & 9.81 &  $m/s^2$  \\\hline
%		$\phi_{min}$ & -0.6981 &  $rad/s$ & $\phi_{max}$ & 0.6981 &  $rad/s$  \\\hline
%		$v_{min}$ & 0 &  $m/s$ & $v_{max}$ & 5 &  $m/s$  \\\hline
%		$u_{1, min}$ & -0.3491 &  $rad/s^2$ & $u_{1,max}$     & 0.3491      & $rad/s^2$ \\\hline $u_{2, min}$ & $-0.3Mg$ &  $m/s^2$ &$u_{2,max}$     & $0.3Mg$      & $m/s^2$  \\ \hline
%	\end{tabular}
%\end{table}

 We also consider the case of implementing the safety constraint (\ref{eqn:safety2}) with a standard HOCBF ($m = 2$, i.e., Eqn. (\ref{eqn:hocbfconstr1})) to make a comparison between the original HOCBF and the HOCBFs proposed in this paper. The simulation trajectories for all cases are shown in Fig. \ref{fig:traj}.

\begin{figure}[htbp]
	\vspace{-4mm}
	\centering
	\subfigure[control input $u_1$]{
		\begin{minipage}[t]{0.24\textwidth}
			\centering
			\includegraphics[width=\textwidth]{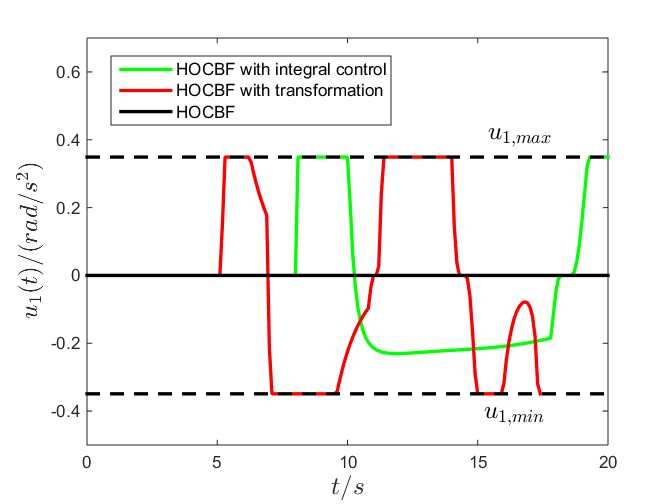}
			%\caption{}
			\label{fig:u1}
		\end{minipage}%
	}%
	\subfigure[control input $u_2/M$]{
		\begin{minipage}[t]{0.24\textwidth}
			\centering
			\includegraphics[width=\textwidth]{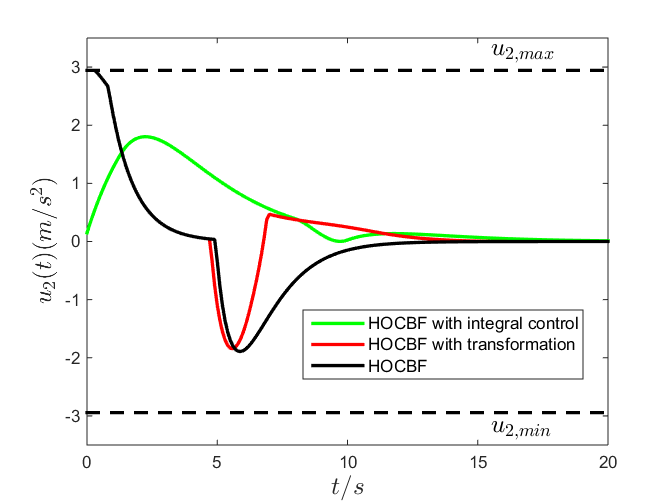}
			%\caption{}
			\label{fig:u2}
		\end{minipage}%
	}%	
	\centering
	\caption{Angular and linear accelerations under different forms of HOCBFs.}\label{fig:av}
	\vspace{-1mm}
\end{figure}

%\begin{figure}[thpb]
%	\centering
%	\includegraphics[scale=0.5]{u1_new}
%	\caption{Angular acceleration profiles (control input $u_1$) under different forms of HOCBFs.}	
%	\label{fig:u1}
%\end{figure}
%\begin{figure}[thpb]
%	\centering
%	\includegraphics[scale=0.5]{u2_new}
%	\caption{Linear acceleration profiles (control input $u_2/M$) under different forms of HOCBFs.}	
%	\label{fig:u2}
%\end{figure}

As shown in Fig. \ref{fig:traj}, if we implement the safety constraint (\ref{eqn:safety2}) with a standard HOCBF, only the control input $u_2$ shows up in the HOCBF constraint (\ref{eqn:hocbfconstr1}). In other words, the robot can only use deceleration to avoid the obstacle, and thus it cannot get to the destination. This is also demonstrated in its control profile in Fig. \ref{fig:u1}  with $u_1(t) = 0, \forall t\in[t_0, t_f]$ (black solid line). 

In the iHOCBF, the relative degree of the HOCBF is higher than the one in the transformation approach. Therefore, we need to pay more effort in the definition of a iHOCBF in terms of the QP feasibility as there are more class $\mathcal{K}$ functions involved. The QP feasibilities of the two proposed approaches are better than the one in the classical HOCBF method whose control $u_1$ is always (constrained to) 0 (the solid black line in Fig. \ref{fig:u1}).

\section{Conclusion}
\label{sec:conc}
We propose two different approaches for high order control barrier functions that work for systems with multiple control inputs. The resulting HOCBFs improve the system performance, as well as improve the problem feasibility. Simulation results on a unicycle model demonstrate the performance and the effectiveness of the proposed approaches. Future work will focus on feasibility analysis and comparison for the proposed approaches under tight control bounds.

\bibliographystyle{IEEEtran}
\bibliography{MCBF}

\end{document}